\documentclass[aps,prl,twocolumn,superscriptaddress,showpacs]{revtex4-2}
\usepackage{graphicx}
\usepackage{dcolumn}
\usepackage{bm}
\usepackage{subfigure}
\usepackage{amsmath}
\usepackage{amssymb}
\usepackage{hyperref}
\usepackage{braket}
\usepackage{siunitx}
\usepackage{xcolor}

\usepackage{verbatim}

\newcommand{\bi}{\boldsymbol{i}}
\newcommand{\bj}{\boldsymbol{j}}

\newcommand{\HC}{\mathrm{h.c.}}
 
\newcommand{\hc}{\hat{c}}
\newcommand{\hcd}{\hat{c}^{\dagger}}

\newcommand{\avs}{$A$V$_3$Sb$_5$}

\begin{abstract}
The \avs~($A=$ K, Rb, Cs) family of Kagome metals hosts unconventional charge density wave (CDW) order whose nature is still an open puzzle. Accumulated evidences point to a time-reversal symmetry breaking (TRSB) orbital flux phase that carries loop currents. Such an order may support anomalous Hall effect. However, the polar Kerr effect measurements that probe the a.c.~anomalous Hall conductivity seem to have yielded contradictory results. We first argue on symmetry grounds that some previously proposed orbital flux order, most notably the one with Star-of-David distortion, shall not give rise to anomalous Hall or polar Kerr effects. We further take the tri-hexagonal orbital flux phase as an exemplary Kagome flux order that does exhibit anomalous Hall response, and show that the Kerr rotation angle at two relevant experimental optical frequencies generally reaches microradians to sub-miliradians levels. A particularly sharp resonance enhancement is observed at around $\hbar \omega =1$ eV, suggesting exceedingly large Kerr rotation at the corresponding probing frequencies not yet accessed by previous experiments. Our study provides important guidance to the interpretation of Kerr measurements on the CDW phase of \avs. In particular, we highlight the important fact that absence of Kerr signal cannot be equated with the absence of TRSB CDW order in \avs. 
\end{abstract}

\begin{document}
	\title{Constraints on the orbital flux phase in \avs~from polar Kerr effect}
	\author{Hao-Tian Liu}
		\address{Shenzhen Institute for Quantum Science and Engineering, Southern University of Science and Technology, Shenzhen 518055, Guangdong, China}
	\address{Beijing National Laboratory for Condensed Matter Physics and Institute of Physics, Chinese Academy of Sciences, Beijing 100190, China}
	\address{University of Chinese Academy of Sciences, Beijing 100049, China}
	\author{Junkang Huang}
	\address{Guangdong Provincial Key Laboratory of Quantum Engineering and Quantum Materials,
School of Physics, Guangdong-Hong Kong Joint Laboratory of Quantum Matter,
and Frontier Research Institute for Physics, South China Normal University, Guangzhou 510006, China}
	\author{Tao Zhou}
	\address{Guangdong Provincial Key Laboratory of Quantum Engineering and Quantum Materials,
School of Physics, Guangdong-Hong Kong Joint Laboratory of Quantum Matter,
and Frontier Research Institute for Physics, South China Normal University, Guangzhou 510006, China}
	\author{Wen Huang}
	\email{huangw3@sustech.edu.cn}
	\address{Shenzhen Institute for Quantum Science and Engineering, Southern University of Science and Technology, Shenzhen 518055, Guangdong, China}
	\address{International Quantum Academy, Shenzhen 518048, China}
	\address{Guangdong Provincial Key Laboratory of Quantum Science and Engineering, Southern University of Science and Technology, Shenzhen 518055, China}
	\maketitle
	
	\setlength\arraycolsep{2pt}

\section{Introduction}
Kagome materials whose active electronic degrees of freedom lie on the underlying Kagome lattice network offer a fertile ground for studying the intricate interplay between quantum many-body correlations, band topology and magnetic geometric frustration~\cite{YJX2022, WY2023,JK2023,SDW2024}. The \avs~($A=$ K, Rb, Cs) family of Kagome metals is a case in point~\cite{BRO2019}. While these compounds do not display any sign of local spin moments~\cite{BRO2019,EMK2021}, they still constitute a model material system which hosts a wide variety of novel physics. In particular, their band structure features a rare combination of nontrivial band topology, Fermi surface nesting and van Hove singularity at the Fermi energy~\cite{BRO2020,BRO2021,YQ2021}. Exactly how electron correlations in this setting promote different electronic instabilities is a matter of considerable scientific interest~\cite{YSL2012,WangWS2013,Kiesel2013}.

At lowest temperatures, all three \avs~compounds develop superconducting pairing~\cite{BRO2020,BRO2021,YQ2021}. Well above the superconducting transition, some form of charge density wave (CDW) order forms at $T_\text{CDW} = $ 78 K, 103 K, 94 K for the $A=$ K, Rb, Cs compounds, respectively~\cite{BRO2020,BRO2021,YQ2021}. The CDW order exhibits an in-plane $2\times 2$ spatial modulation~\cite{ZH2021,JYX2021,NY2021,WZ2021}, with possible three-dimensional structure~\cite{LZ2021,LH2021,HY2022,QS2022,Frassineti:23,HZ2023}.
The CDW phase in \avs~materials is crucial for understanding their superconducting characteristics and holds the key to potentially inducing topological phase transitions~\cite{rwang2023}. This phase represents the forefront of research, shedding light on the complex interactions between topological properties, symmetry breaking, and the unique electronic behavior in these pioneering superconducting systems. In recent years, the CDW phase in \avs~has garnered widespread interest due to its significant potential to influence the development of quantum technologies and to enhance our foundational understanding of electronic behavior in complex lattice structures.

The nature of the CDW phase is still a subject of ongoing investigation, as indicated by recent studies~\cite{JK2023,SDW2024}. Some of its highly unconventional characters have been revealed. First, a nematic order appears to onset at a temperature intermediate between $T_\text{CDW}$ and the superconducting transition, lowering the six-fold rotation symmetry of the underlying Kagome lattice to two-fold~\cite{XY2021,LH2022,NL2022}. Second, the charge order breaks time-reversal invariance, as suggested in $\mu$SR studies~\cite{YL2021,CM2022,RK2022}, as well as in transport measurements of anomalous Hall effect~\cite{YSY2020,YFH2021,WL2023}. The time-reversal symmetry breaking (TRSB) is further implicated by the observation of field-switchable electronic chirality in tunneling spectroscopic~\cite{JYX2021,NY2021} and electro-magneto chiral anisotropy~\cite{GC2022} measurements. At this stage, the leading TRSB CDW candidate in \avs~is the so-called orbital flux phase, which is characterized by spontaneous bond currents that form patterns of loop currents. Multiple $2\times 2$ orbital flux phases have been proposed~\cite{FX2021a, Denner:21,TH2021,ParkT:21,LYP2021,HS2023}. Among them, two receive the most attention~\cite{JK2023,SDW2024}: Star-of-David (SoD) and tri-hexagonal (TrH) patterns, the latter of which is also referred to as Inverse Star-of-David. Gaining a comprehensive understanding of these orbital flux phases is essential for deciphering the TRSB CDW phase in \avs, challenging prevailing assumptions, and offering novel insights into the superconducting properties and topological features of these captivating materials.

Another would-be strong evidence of TRSB is the polar Kerr effect~\cite{AK2015}, wherein a polarized probing light shone vertically on a sample is reflected with a rotated polarization angle. The majority of Kerr effect measurements have been conducted on CsV$_3$Sb$_5$, and they have yielded conflicting results. Two studies performed at wavelength 800 nm ($\hbar \omega = 1.55$ eV) reported low-temperature Kerr rotations as large as tens of microradians to sub-milliradians~\cite{WQ2022,XY2022}. Four other measurements at 1550 nm wavelength ($\hbar\omega = 0.8$ eV) observed considerably smaller Kerr rotation, with one obtaining 2 $\mu$rad \cite{HYJ2022} and the other three getting negligible values not exceeding 0.03 $\mu$rad \cite{DRS2023,CF2023,WJY2024}. The latter four studies used the Sagnac interferometer setup supposedly more sensitive to TRSB. While the large discrepancy between measurements at the two wavelengths may be attributed to different resonance enhancement, Ref.~\onlinecite{CF2023} (1550 nm) also revealed a milliradian-size signal which curiously does not reverse sign with opposite training fields. As a consequence, it cannot be ruled out that the large signal at the two wavelengths may both have other origin, instead of a TRSB orbital flux order. 

We note that, even if Kerr rotation is not observed, TRSB cannot be definitively ruled out. For example, Kerr effect is absent in TRSB antiferromagnet-like loop-current orders wherein the flow of current in the loops alternates in sign (see {\it e.g.}~Ref.~\onlinecite{Tewari:08}). Nevertheless, Kagome lattice can in principle host orbital flux phases containing more complicated bond current textures than solely closed current loops. In this case, the orbital currents cannot be directly classified as ferro- or antiferro-type of loop currents, and more systematic analyses may be required. One goal of this study is to explore the implication of the presence or absence of the Kerr effect on the nature of the orbital flux phase in \avs. We revisit the symmetry constraint on the anomalous Hall effect (AHE), which constitutes the microscopic origin of the polar Kerr effect. We shall see that, the Kerr effect is expected in the TrH phase but not in the SoD phase. 

Under the assumption that Kerr effect is indeed present in \avs, we further aim to address whether the discrepancy between the 800-nm and 1550-nm measurements can be explained. Taking the TrH phase as an example, we evaluate the anomalous Hall conductivity and the corresponding Kerr rotation angle, and find that the Kerr angle at both wavelengths generally varies between microradian and sub-milliradian levels. The exact values depend rather sensitively on the microscopic details of the loop currents, and fine tuning may be needed to reach face-value agreement with existing measurements at the two wavelengths~\cite{WQ2022,XY2022,HYJ2022,DRS2023,CF2023,WJY2024}. Interestingly, a strong resonance enhancement originating from pronounced interband Berry curvature involving the flatbands is seen at around 1240 nm ($\hbar \omega = 1$ eV). This implies large Kerr rotation around this probing wavelength, suggesting a potential future direction to pursue. 
	
	\begin{figure}
	\includegraphics[width=8.5cm]{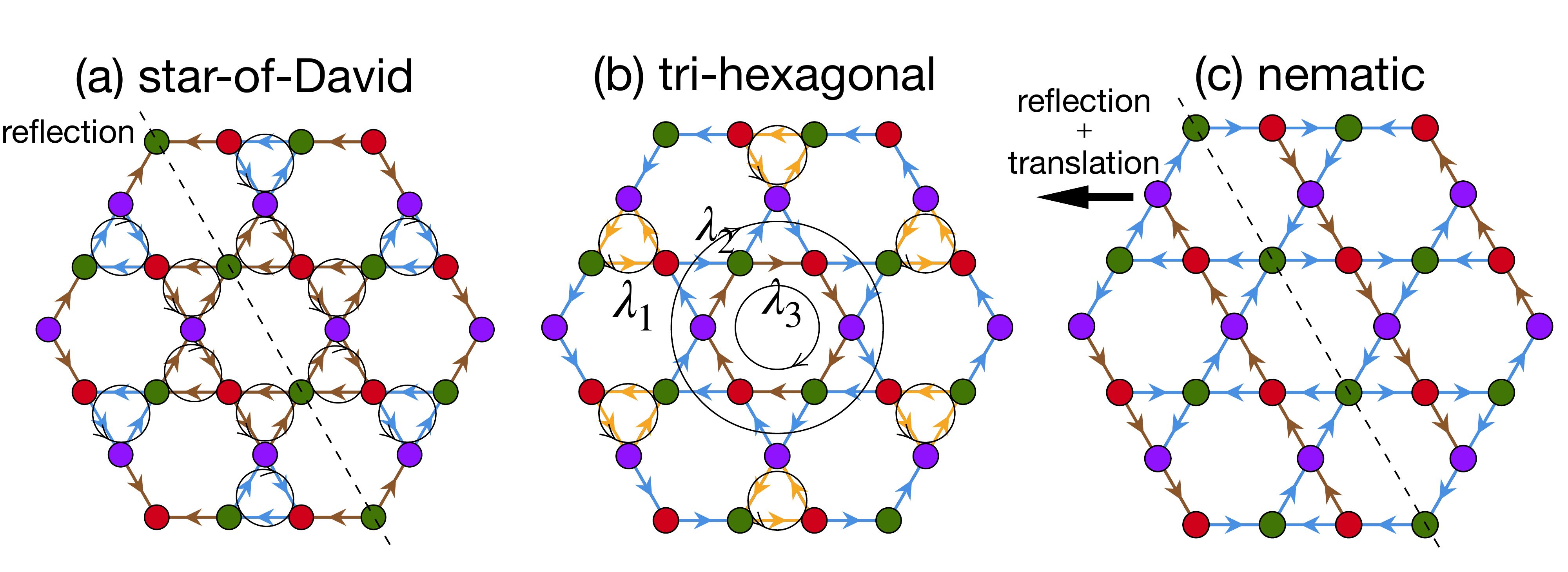}
	\caption{Schematics of three representative orbital flux phases on the Vanadium Kagome lattice in \avs~literature. The arrows indicate the flow direction of the bond currents, and the bond colors encode the magnitude of the bond currents. The thin black circles with arrows sketch the overall flow direction of the current loops. The dashed lines in (a) and (c) represent the mirror reflection axis, with mirror planes normal to the Kagome lattice. }
	\label{fig:loopcurrent}
	\end{figure}

\section{ Symmetry arguments} 

While TRSB is a prerequisite, it alone does not constitute a sufficient condition for the occurrence of the AHE in linear response. A TRSB electronic order in two-dimension that supports AHE must also break all reflection symmetries with mirror planes perpendicular to the system plane, or, the product of such vertical-plane mirror symmetry and an in-plane translation symmetry~\cite{WYX2014, WC2016}. Pictorially, this is similar to having a net out-of-plane magnetization originating from either orbital currents or spin polarization. The resultant AHE gives rise to the polar Kerr effect in the backscattering geometry, wherein the polarized light is normally incident on the sample. 

To understand the above symmetry constraints for orbital flux phases, let's focus on the Hall conductivity that follows from the Kubo formula,
	\begin{align}\label{hall0}
	\sigma_{H}(\omega)=\frac{i}{2\omega}[\pi_{xy}(\omega+i0^+)-\pi_{yx}(\omega+i0^+)],
	\end{align}
	where $\pi_{\mu\nu}$ is the current-current correlation function, which is defined as follows,
	\begin{align}
	\pi_{\mu\nu}(i\nu_m)=\frac{1}{\beta A} \sum_{\mathbf{k},\omega_n} \operatorname{tr}[\hat{V}_{\mu\mathbf{k}} G(\mathbf{k}, i\omega_n) \hat{V}_{\nu\mathbf{k}} G(\mathbf{k}, i\omega_n + i\nu_m)],
	\end{align}where $A$ represents the area of the system, $G(\mathbf{k},i\omega_n)=(i\omega_n-H_\mathbf{k})^{-1}$ is the Green's function where $H_{\mathbf{k}}$ is the system Hamiltonian, $\hat{V}_{\mu\mathbf{k}}=\partial_{k_{\mu}} H_{\mathbf{k}} \equiv \partial_{\mu} H_{\mathbf{k}} $ stands for the $\mu$-th component of the velocity operator, $\omega_n$ and $\nu_m$ denote, respectively, the fermionic and bosonic Matsubara frequencies, and $\omega$ is the real frequency. Notably, the Hall effect probed in the d.c.~limit is related to $\text{Re}[\sigma_H(\omega=0)]$, whereas the a.c.~Hall response, as manifested for example in the polar Kerr effect, would include contributions from both the real and imaginary parts of finite-frequency $\sigma_H$. 
	
	To facilitate further analysis, it is informative to turn to the spectral representation of the Green's function,
	\begin{align}
	G(\mathbf{k}, i\omega_n) = (i\omega_n - H_{\mathbf{k}})^{-1} = \sum_{a} \frac{|\psi_{a\mathbf{k}}\rangle \langle \psi_{a\mathbf{k}}|}{i\omega_n - \epsilon_{a\mathbf{k}}},
	\end{align}
	where $|\psi_{a\mathbf{k}}\rangle$ is the cell-periodic part of the Bloch wavefunction of band-$a$, whose band dispersion is given by $\epsilon_{a\mathbf{k}}$. The Hall conductivity then follows as,
	\begin{align}
	\sigma_{H}(\omega) &= \frac{i}{2A\omega} \sum_{\mathbf{k}, a, b} \frac{f(\epsilon_{b\mathbf{k}}) - f(\epsilon_{a\mathbf{k}})}{\omega - \epsilon_{a\mathbf{k}} + \epsilon_{b\mathbf{k}} + i\eta} \Gamma^{ab}_{xy}(\mathbf{k}),\label{eq:hall}
	\end{align}
	where $f(E)$ is the Fermi-Dirac distribution function, and 	
	\begin{eqnarray}
	\Gamma^{ab}_{xy}(\mathbf{k}) &=& \langle \psi_{a\mathbf{k}} |\hat{V}_{x \mathbf{k}} |\psi_{b\mathbf{k}}  \rangle \langle \psi_{b\mathbf{k}} |\hat{V}_{y \mathbf{k}}| \psi_{a\mathbf{k}}  \rangle - (x \leftrightarrow y) \nonumber \\
	&=&  -(\epsilon_{a\mathbf{k}}-\epsilon_{b\mathbf{k}})^2 \mathcal{B}^{ab}_{xy}(\mathbf{k}) \,.
	\end{eqnarray}
Here, $\mathcal{B}^{ab}_{xy}(\mathbf{k})$ is the interband Berry curvature given by:
	\begin{align}\label{interbandberrycurvature}
 i\left(\langle \partial_{x} \psi_{a\mathbf{k}} | \psi_{b\mathbf{k}} \rangle \langle \psi_{b\mathbf{k}} | \partial_{y} \psi_{a\mathbf{k}} \rangle - \langle \partial_{y} \psi_{a\mathbf{k}} | \psi_{b\mathbf{k}} \rangle \langle \psi_{b\mathbf{k}} | \partial_{x} \psi_{a\mathbf{k}} \rangle \right)
	\end{align}
We thus see that whether Hall response can arise is determined by the momentum-space distribution of $\mathcal{B}^{ab}_{xy}(\mathbf{k})$. 
	
	Consider first a two-dimensional system exhibiting a product of mirror and translation symmetries, which includes a vertical-plane mirror reflection $\mathcal{M}_{y}$ that sends coordinate $(x,y)$ to $(x,-y)$ and a translation of half the lattice constant along the $x$-axis, $\mathcal{T}_{x,\frac{a}{2}}$. The combined operation $\mathfrak{g} = \mathcal{M}_y\mathcal{T}_{x,\frac{a}{2}} = \mathcal{T}_{x,\frac{a}{2}} \mathcal{M}_y$ transforms a wavefunction $\psi(x, y)$ in real-space coordinates as $\mathfrak{g}\psi(x, y) = \psi(x + \frac{a}{2}, -y)$. In momentum space, this transformation yields $\mathfrak{g}\psi_{k_x, k_y} = e^{i k_x a/2} \psi_{k_x, -k_y}$, one then has $\mathcal{B}^{ab}_{xy}(k_x, k_y) = -\mathcal{B}^{ab}_{xy}(k_x, -k_y)$~\cite{supplementary1}. As a consequence, the Hall conductivity in Eq.~\eqref{eq:hall} vanishes after integrating over $\mathbf{k}$! The above analysis can be easily generalized to the simpler scenarios where the system preserves pure vertical-plane mirror symmetries. 
	

	Previously, the $2\times 2$ orbital flux phases on the Kagome lattice have been extensively classified~\cite{FX2021b, MC2022,GW2023}. However, we do not seek to assess the occurrence of AHE on the basis of the previous classifications. Instead, we shall focus on the flux phases most frequently discussed in the \avs~context, namely, the TrH and the SoD phases. Figure \ref{fig:loopcurrent} sketches the patterns of $2\times 2$ bond currents on the Kagome lattice for both of these two phases. As one can check, the TrH phase has neither vertical mirror symmetry nor its product with translation, and is therefore expected to generate AHE. By contrast, the SoD phase preserves mirror symmetries associated with $\mathcal{M}_y$, indicating vanishing AHE. Note that this phase also preserves $C_3$ and $C_6T$ symmetries, where $T$ denotes time-reversal operation. Also presented in Figure \ref{fig:loopcurrent} is an example of a nematic orbital flux phase~\cite{Denner:21,FG2023,JiangHM:23}, a descendant of the TrH phase that breaks all vertical-plane mirror symmetries but respects the product of a mirror and a translation of half a lattice constant of the $2\times 2$ enlarged CDW unitcell. Hence, this state does not support AHE either. The lack of Hall response in the latter two can also be inferred by noting the vanishing of net orbital flux due to an exact cancellation among opposite loop currents within each $2\times 2$ unit cell. The above conclusions will be separately verified by our numerical calculations later. It is worth cautioning that we have thus far assumed ideal CDW patterns without structural chirality that breaks all vertical mirror symmetries as indicated in some surface scanning probes~\cite{JYX2021, NY2021,XingY2024}. In principle, SoD and nematic orbital flux phases further distorted with such structural chirality shall also exhibit Hall and Kerr effects. However, since it is not certain whether the observed structural chirality is a surface-only effect, we shall not delve into these scenarios in the present study. 
	
	The above analyses can also be generalized to scenarios with three-dimensional modulation of the CDW order. Experimentally, there have been hints of alternate stacking of TrH and SoD layers~\cite{HY2022}, or a $\pi$-shifted stacking of the TrH order~\cite{Frassineti:23}. Both of these cases contain TrH layers and therefore will exhibit AHE. On the other hand, a $\pi$-shifted stacking of the SoD order will not show Hall response, as it still respects the mirror symmetry of the single-layer limit. 
	

    \begin{figure}
    \includegraphics[width=8.8cm]{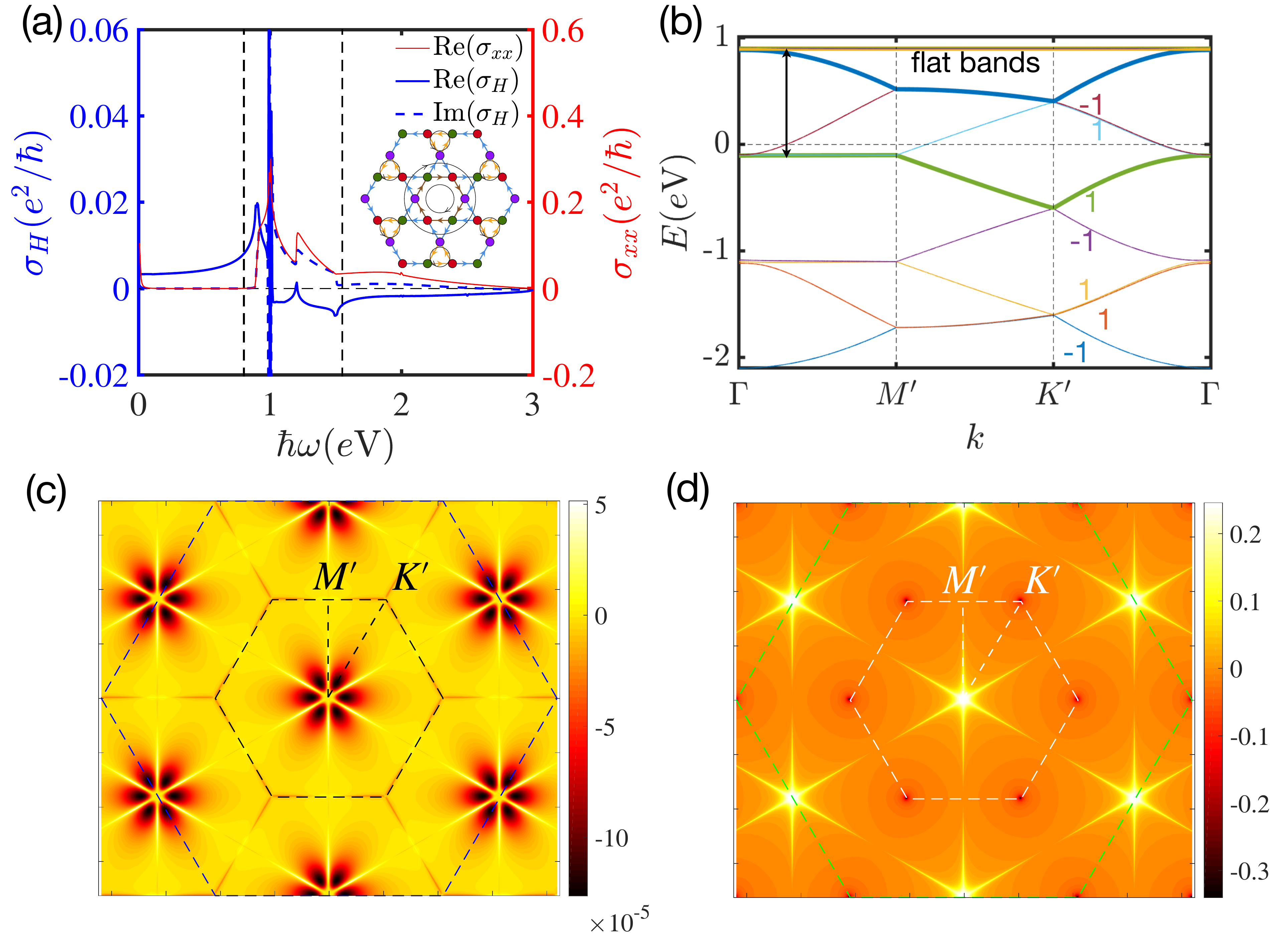}
    	\caption{Zero-temperature AHE in the TrH phase with $\{t,\mu,\lambda_1, \lambda_2,\lambda_3\} = \{0.5, 0.1,-0.005,-0.005,0.01\}$eV. (a) Real and imaginary parts of the anomalous Hall conductivity $\sigma_H$ and the real part of the longitudinal conductivity $\sigma_{xx}$. The two vertical dashed lines indicate $\hbar\omega=$ 0.8 eV and 1.55 eV, which correspond to the existing experimental optical wavelengths 1550 nm and 800 nm, respectively. (b) Band structure and Chern number of the individual bands. (c) and (d) Momentum-space distribution of the interband Berry curvature $\mathcal{B}^{ab}_{xy}$ with $\hbar\omega=1$ eV and between bands whose dispersions are highlighted by thick curves in (b), namely, (c) between the thick green and thick blue bands, and (d) between the thick green and thick yellow (flat) bands. }
    	\label{fig:TrHresults}
    \end{figure}

\section{Model calculations} 

We construct the tight-binding model of different $2\times 2$ orbital flux phases on the Kagome lattice, assuming that the band structure around the Fermi level is dominated by a single Vanadium $d$-orbital residing on each lattice site. As spin order has not been reported in the \avs~family, we shall use a spinless model for simplicity, keeping in mind that the final numerical results shall be multiplied by two to account for the two spin degrees of freedom. In real space, the Hamiltonian is given by~\cite{footnote},
	\begin{align}\label{fluxHamiltonian}
	H&=-t\sum_{\braket{\bi, \bj}}(\hcd_{\bi}\hc_{\bj}+\HC)-\mu\sum_{\bi}\hcd_{\bi}\hc_{\bi}\nonumber \\
	&-i\sum_{\braket{\bi,\bj}}\lambda_{\bi \bj}(\hcd_{\bi}\hc_{\bj}-\HC) \,.
	\end{align}
	Here, $t$ represents the nearest-neighbor hopping amplitude, $\lambda_{\bi \bj}$ designates the bond current flowing from $\bj$th site to $\bi$th site, $\mu$ is the chemical potential, and $\braket{\bi \bj}$ denotes nearest neighbors. With $t\simeq 0.5$ eV and proper filling fraction, and without the last term which describes the orbital flux order, the Hamiltonian is able to capture well both the normal-state band structure and the Fermi surface geometry of the low-energy Vanadium-dominated bands obtained in first principle studies~\cite{GuY:21}. The sign of $\lambda_{\bi\bj}$ follows the arrow direction depicted for the individual flux phases in Fig.~\ref{fig:loopcurrent}, with the convention that a positive $\lambda_{\bi\bj}$ indicates a bond current flowing from $\bi$ to $\bj$, and vice versa. The $2\times 2$ flux phase quadruple the unit cell, resulting in twelve lattice sites in the enlarged unit cell. 
	
	We transform the above Hamiltonian \eqref{fluxHamiltonian} into momentum-space formulation, after which the optical Hall conductivity can be conveniently computed according to Eqns.~\eqref{hall0} or \eqref{eq:hall}. Consistent with the above symmetry arguments, our numerical calculations confirm the vanishing of the Hall conductivity in the SoD and nematic orbital flux phases. 
	
	In the following, we focus on the TrH phase. According to the sketch in Fig.~\ref{fig:loopcurrent}, each set of bonds designated the same color form a closed current loop. The bond current amplitudes on the three loops are given by $\{\lambda_1, \lambda_2,\lambda_3\}$. Except for fine-tuned set of $\lambda_i$'s, these current loops in general exhibit a finite net orbital flux (magnetization) in each $2\times 2$ cell. The resultant model is a Chern metal, and the Chern numbers of the individual bands are labeled in Fig.~\ref{fig:TrHresults} (b) alongside the band structure. Figure \ref{fig:TrHresults} (a) shows the real and imaginary parts of $\sigma_H$ obtained at zero temperature for a representative set of parameters close to the CDW gap of order $10$ meV inferred from tunneling and optical spectroscopic measurements~\cite{JYX2021,LZ2021,KangM:22}. According to \eqref{eq:hall}, $\text{Im}[\sigma_H]$ is related to the interband transition spectrum and is given by,
	\begin{equation}
	\frac{1}{2A}\sum_{\mathbf{k},a,b} [f(\epsilon_{a\mathbf{k}}) - f(\epsilon_{b\mathbf{k}})] (\epsilon_{a\mathbf{k}} - \epsilon_{b\mathbf{k}}) \mathcal{B}^{ab}_{xy}(\mathbf{k})\delta(\hbar\omega -\epsilon_{a\mathbf{k}} +\epsilon_{b\mathbf{k}}) \,.
	\end{equation}
	
	As one can see from Fig.~\ref{fig:TrHresults} (a), $\text{Im}[\sigma_H]$ is characterized by broad resonance features between around 0.9 eV and 1.5 eV. This resonance window varies with $\lambda_i$'s, but generally falls within the resonance of the longitudinal optical conductivity $\text{Re}{[\sigma_{xx}]}$ that spans a broader frequency range (Fig.~\ref{fig:TrHresults} (a)). Here, $\sigma_{xx}$ is related to the longitudinal current-current correlation and is evaluated according to,
	\begin{align}
		\sigma_{xx}(\omega) & =  \frac{i}{A\omega} \sum_{\mathbf{k}, a, b} \frac{|\braket{\psi_{a\mathbf{k}}|\hat{V}_{x\mathbf{k}}|\psi_{b\mathbf{k}}}|^2[f(\epsilon_{b\mathbf{k}}) - f(\epsilon_{a\mathbf{k}})]}{\omega - \epsilon_{a\mathbf{k}} + \epsilon_{b\mathbf{k}} + i\eta}\,.
	\end{align}
	 Besides the difference in the resonance range, the lineshape of the two types of conductivities also differ somewhat. These differences suggests that the Hall response may be associated more prominently with some interband optical transitions than the others. 
	
	To gain further insight, we plot in Fig.~\ref{fig:TrHresults} (c) and (d) the momentum-space distribution of the interband Berry curvature $\mathcal{B}^{ab}_{xy}$ for two different sets of bands. In most cases, we see that $\mathcal{B}^{ab}_{xy}$ is sharply peaked in very restricted areas in the Brillouin zone. We thus arrive at an interesting observation that, the CDW-induced band topology is most strongly featured by a small fraction of the Bloch electrons in each band, and this in turn limits the type of interband transitions capable of generating Hall response. Moreover, $\mathcal{B}^{ab}_{xy}$ involving one of the flatbands and the band marked by thick green band dispersion is particularly large. The corresponding interband transitions (symbolized by the double arrow in Fig.~\ref{fig:TrHresults} (b)) are responsible for the sharp resonance at $\hbar\omega \simeq 1$ eV. Notably, this feature is seen in a broad parameter range of bond currents we have considered.
	
	In comparison, $\mathcal{B}^{ab}_{xy}$ for SoD and nematic phases vanishes everywhere in the Brillouin zone (not shown in figure). This in fact is ensured by symmetry, {\it e.g.}, for the SoD phase a combination of $\mathcal{M}_y$, $C_3$ and $C_6T$ symmetries.

		\begin{figure}
	    \includegraphics[width=8.7cm]{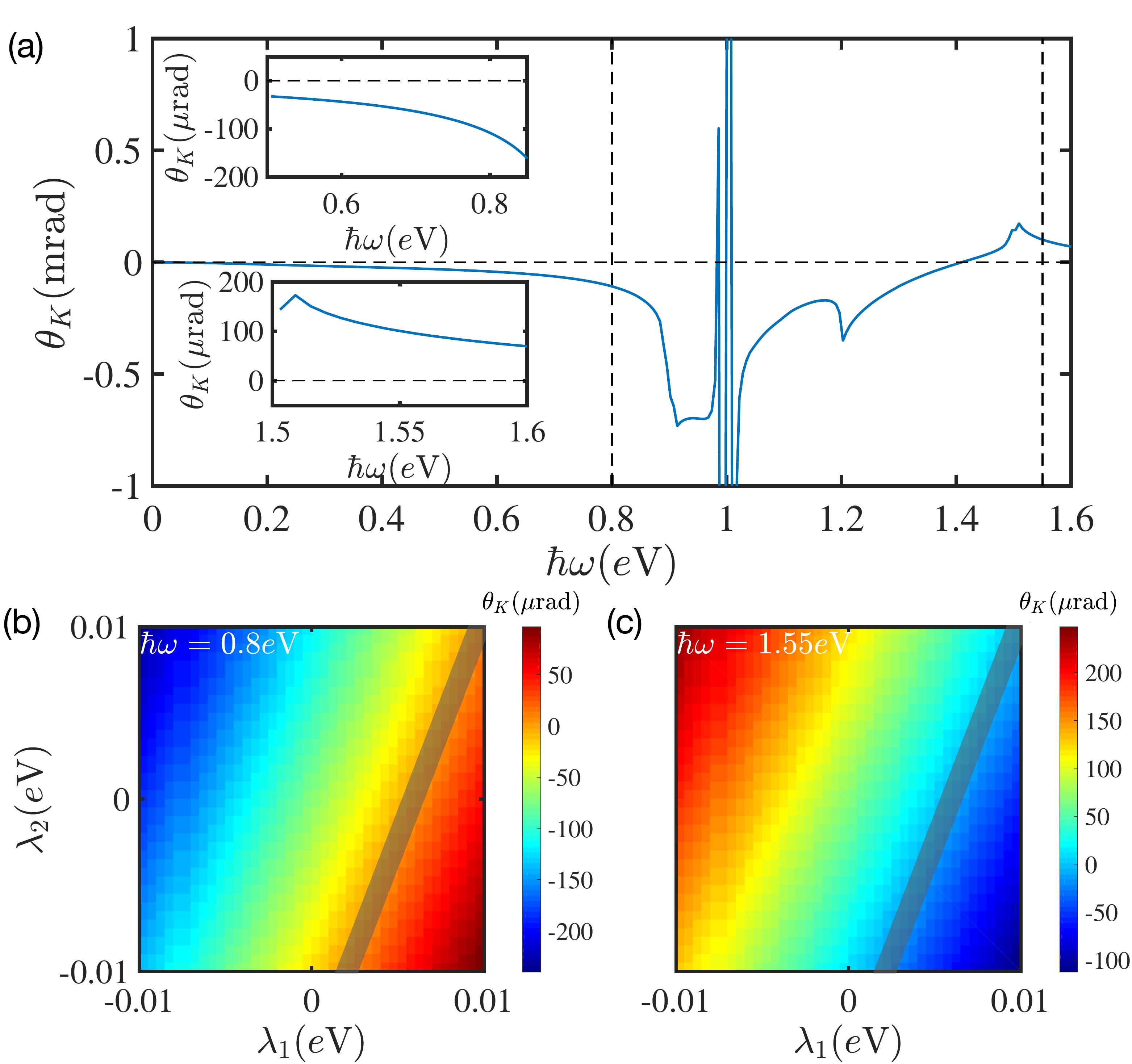}
	    	\caption{Zero-temperature Kerr rotation angle in the TrH phase. (a) Estimated Kerr angle with $\{t,\mu,\lambda_1, \lambda_2,\lambda_3\} = \{0.5, 0.1,-0.005,-0.005,0.01\}$eV. The insets show zoom-in view around the two frequencies already accessed by Kerr effect studies, $\hbar\omega=0.8$ and 1.55 eV. (b) and (c) Variation of the Kerr angle at the two experimental frequencies as a function of $\lambda_1$ and $\lambda_2$, with $\lambda_3 \equiv 0.01$ eV. The shaded area (same in both subfigures) roughly corresponds to $|\theta_K(\hbar\omega=0.8~\text{eV})| \leq 5$ $\mu$rad.}
	    	\label{fig:Kerr}
	    \end{figure}

\section{Kerr angle in tri-hexagonal phase}	
In the backscattering geometry, the Kerr rotation angle is related to the Hall conductivity as~\cite{PA1955}
	\begin{align}\label{eq:KerrEq}
	\theta_{K}(\omega) = \frac{4\pi}{\omega d}\operatorname{Im}\bigg[\frac{\sigma_{H}(\omega)}{n(\omega)[n(\omega)^2-1]}\bigg] \,,
	\end{align}
	where $d$ is the interlayer spacing in \avs, and $n(\omega)$ denotes the frequency-dependent refraction index. Using the same set of parameters as in Fig.~\ref{fig:TrHresults}, Fig.~\ref{fig:Kerr} (a) presents the estimated zero-temperature Kerr angle as a function of the probing frequency. Details of the computation are provided in the Supplementary~\cite{supplementary1}. Since $\text{Re}[\sigma_H(\omega)]$ also contributes in \eqref{eq:KerrEq}, the Kerr angle does not diminish beyond the resonance of $\text{Im}[\sigma_H]$. Existing Kerr measurements on \avs~\cite{WQ2022,XY2022,HYJ2022,DRS2023,CF2023,WJY2024} were performed at fixed optical frequencies $\hbar\omega=0.8$ eV (1550 nm) and 1.55 eV (800 nm). In the present calculation, both frequencies see significant Kerr rotation angle of the order of 100 $\mu$rad (see insets of Fig.~\ref{fig:Kerr} (a)). At face value, this would seem consistent with the 800-nm measurements~\cite{WQ2022,XY2022} and disagree with the 1550-nm ones~\cite{HYJ2022,DRS2023,CF2023,WJY2024}. However, we note possible sizable disorder effects~\cite{Nakazawa2024}, and more importantly the strong sensitivity to the microscopic details of the bond current configuration. Figure \ref{fig:Kerr} (b) and (c) show representative variation of $\theta_K$ at the two frequencies as a function of the bond currents. With one of the bond currents, $\lambda_3$, held fixed at 10 meV, $\theta_K$ generally varies on the scale of microradians to sub-miliradians for $\lambda_1$ and $\lambda_2$ in the range of $\pm 10$ meV. Furthermore, $\theta_K$ exhibits a linear relation separately with each of the bond currents, and can change sign within the range of parameters considered. Highlighted in gray shading is the regime with $|\theta_K(\hbar\omega=0.8~\text{eV})| \leq 5$ $\mu$rad. This regime of parameters leads to similar estimates of the Kerr angle at $\hbar\omega=1.55~\text{eV}$ (Fig.~\ref{fig:Kerr} (c)), in contrast to the 800-nm measurements~\cite{WQ2022,XY2022}. One would thus need extra fine tuning in order to reconcile the Kerr measurements done at the two wavelengths. Nonetheless, we also take note of the possible difference caused by the different experimental setup as discussed in Refs.~\onlinecite{DRS2023,WJY2024}.
	
	Finally, the Kerr angle at around $\hbar\omega =1$ eV readily reaches the miliradians range, tracking the parameter-insensitive sharp resonance enhancement of the Hall conductivity. Hence, had the \avs~compounds indeed developed the TrH phase, Kerr measurements at around 1240 nm would likely yield the largest outcome. 
	
\section{Summary} 
In an attempt to interpret the multiple polar Kerr effect measurements in the \avs~compounds, we have carried out symmetry analyses and numerical modeling of the AHE for the orbital flux phases on the Kagome lattice. The symmetry argument showed that the previously proposed SoD (Fig.~\ref{fig:loopcurrent} (a)) and nematic orbital flux phases (Fig.~\ref{fig:loopcurrent} (c)) do not support AHE nor polar Kerr effect, as they each preserve certain vertical-plane mirror symmetry or its product with a lattice translation. We further took the TrH orbital flux phase (Fig.~\ref{fig:loopcurrent} (b)) as an exemplary flux order that do exhibit Kerr effect, and evaluated its anomalous Hall conductivity as well as the related Kerr rotation angle. Our results suggest that, for TrH bond current amplitudes relevant to \avs, the low-temperature Kerr angle at existing experimental photon frequencies generally reach microradians to sub-miliradians. In addition, we observed a sharp resonance enhancement at around $\hbar\omega =1$ eV, which is related to interband Berry curvature involving the flatbands. The polar Kerr rotation, if existent, may thus be significantly larger in the regime of optical wavelengths ($\sim 1240$ nm) not yet accessed by previous measurements. Our study provides a basis for interpreting Kerr measurements on the CDW phase of the \avs~compounds. In particular, it serves as an important reminder that, one cannot rule out TRSB CDW order in \avs~simply based on the absence of Kerr signal reported in some recent measurements~\cite{CF2023,DRS2023,WJY2024}.

\section{Acknowledgement} We thank Jun-Feng Dai, Qiye Liu, Xianxin Wu, Luyi Yang, and Jia-Xin Yin for helpful discussions. This work is supported by NSFC under grants No.~12374042 and No.~11904155, the Guangdong Science and Technology Department under Grant 2022A1515011948, a Shenzhen Science and Technology Program (Grant No.~KQTD20200820113010023), and the Guangdong Provincial Key Laboratory under Grant No.~2019B121203002. Computing resources are provided by the Center for Computational Science and Engineering at Southern University of Science and Technology.

            \clearpage
        	\appendix
        	\setcounter{equation}{0}
        	\setcounter{figure}{0}
        	\renewcommand {\theequation} {S\arabic{equation}}
        	\renewcommand {\thefigure} {S\arabic{figure}}
        	
        	
        	\section{Supplementary Materials}
        	\label{S1}        	
        	
        	\textit{Proof of $\mathcal{B}^{mn}_{xy}(k_x, k_y) = -\mathcal{B}^{mn}_{xy}(k_x, -k_y)$ under $\mathcal{M}_y\mathcal{T}_{x,\frac{a}{2}}$ symmetry}--- In this section, we show how the Hall conductivity vanishes for a model with a combined mirror reflection and a half-lattice-constant translation symmetries. As defined in the main text, the combined operation is given by $\mathfrak{g} = \mathcal{M}_y\mathcal{T}_{x,\frac{a}{2}}$, which acts on a Bloch state according to $\mathfrak{g}\psi_{k_x, k_y} = e^{i k_x a/2} \psi_{k_x, -k_y}$. One then has

\begin{widetext}
        	\begin{eqnarray}
 \mathcal{B}^{mn}_{xy}(k_x,k_y) &=& \braket{\partial_{k_x}\psi_{m,k_x,k_y}|\mathfrak{g}^{-1}\mathfrak{g}|\psi_{n,k_x,k_y}}\braket{\psi_{n,k_x,k_y}|\mathfrak{g}^{-1}\mathfrak{g}|\partial_{k_y}\psi_{m,k_x,k_y}}  - (\partial_{k_x} \leftrightarrow \partial_{k_y}) \nonumber \\
  &=& \partial_{k_x}\left[ e^{-ik_xa/2}\langle \psi_{m,k_x,-k_y}|\right] e^{ik_xa/2}|\psi_{n,k_x,-k_y}\rangle \langle \psi_{n,k_x,-k_y}| e^{-ik_xa/2} \partial_{k_y} \left[e^{ik_xa/2}|\psi_{m,k_x,-k_y} \rangle \right]  - (\partial_{k_x} \leftrightarrow \partial_{k_y}) \nonumber \\
  &=&  \braket{\partial_{k_x}\psi_{m,k_x,-k_y}|\psi_{n,k_x,-k_y}}\braket{\psi_{n,k_x,-k_y}|\partial_{k_y}\psi_{m,k_x,-k_y}}  - (\partial_{k_x} \leftrightarrow \partial_{k_y}) \nonumber \\
  &=& -  \mathcal{B}^{mn}_{xy}(k_x,-k_y) \,.
        	\end{eqnarray}
\end{widetext}
In getting the third line we have used the orthogonality condition $\langle \psi_{m,k_x,-k_y}|\psi_{n,k_x,-k_y}\rangle =0$ for any $m \neq n$. Using Eqs.~(4) and (5) in the maintext we then see that the Hall conductivity vanishes after integrating over $\mathbf{k}$. Similar proof can be made for models with $\mathcal{M}_y$ symmetry. 
        	
        	\textit{Calculation of the Kerr angle}---The optical Kerr angle in SI units is give by~\cite{MG2013}
        	\begin{align}\label{kerr_S}
        		\theta_{K}(\omega) = \frac{1}{\epsilon_0 \omega d}\operatorname{Im}\biggl\{\frac{\sigma_H(\omega)}{n(\omega)[n^2(\omega)-1]}\biggr\},
        	\end{align}
        	where $\epsilon_0\simeq \SI[per-mode=symbol]{8.85e-12}{\coulomb\squared\per\joule\per\meter}$ is the vacuum permittivity, $d$ is the interlayer spacing along $c$-axis, and $n(\omega)$ is the frequency-dependent complex refractory index, given by
        	\begin{align}
        		n(\omega) &= \sqrt{\epsilon(\omega)},\\
        		\epsilon(\omega) &= \epsilon_{\infty} + \frac{i}{\omega}\frac{\sigma_{xx}(\omega)}{\epsilon_0}.
        	\end{align}
        	Here $\epsilon(\omega)$ is the permeability tensor of the vanadium sublattice, $\epsilon_{\infty}$ is the background permeability, and $\sigma_{xx}(\omega)$ is the complex optical conductivity. To accurately compute the Kerr angle, it is crucial to extract precise values from experimental data in  $A$V$_3$Sb$_5$ for the quantities $d$, $\epsilon_{\infty}$ and $\sigma_{xx}(\omega)$. 
        	
        	\begin{table}[h]
        	            \centering
        	            \renewcommand{\arraystretch}{1.5}
        	            \setlength{\tabcolsep}{8pt}
        	            \begin{tabular}{|c|c |c|}
        	                \hline
        	              Drude   & $k=1$ & $k=2$ \\
        	                \hline
        	               $\omega_{p,k}^2(\times 10^{8}\mathrm{cm}^{-1})$  & $3.3$ & $3.3$ \\
        	                
        	                $\tau_k^{-1}(\times 10^3\mathrm{cm}^{-1})$ & 0 & $0.5$ \\
        	                \hline
        	                \end{tabular}
        	                \caption{Parameters of the Drude oscillators.}
        	                \label{drude}
        	            \end{table}
        	            
        	            \begin{table}[h]
        	            \centering
        	            \renewcommand{\arraystretch}{1.5}
        	                \setlength{\tabcolsep}{8pt}
        	            \begin{tabular}{|c|c|c|c|}
        	                \hline
        	              Lorentz   & $j=1$ & $j=2$ & $j=3$ \\
        	                \hline
        	                $\omega_{0,j}(\times 10^3\mathrm{cm}^{-1})$ & 5.7 & 9.9 & 1.8 \\
        	                
        	                $\gamma_j(\times 10^3\mathrm{cm}^{-1})$ & 2.5 & 3 & 1.9 \\
        	                
        	                $\omega_{p,j}^2(\times 10^{8}\mathrm{cm}^{-1})$ & 2.7 & 3.8 & 6 \\
        	                \hline
        	                \end{tabular}
        	                \caption{Parameters of Lorentz oscillators.}
        	                \label{lorentz}
        	            \end{table}
        	
\begin{figure}[htbp]
            	\includegraphics[width=6cm]{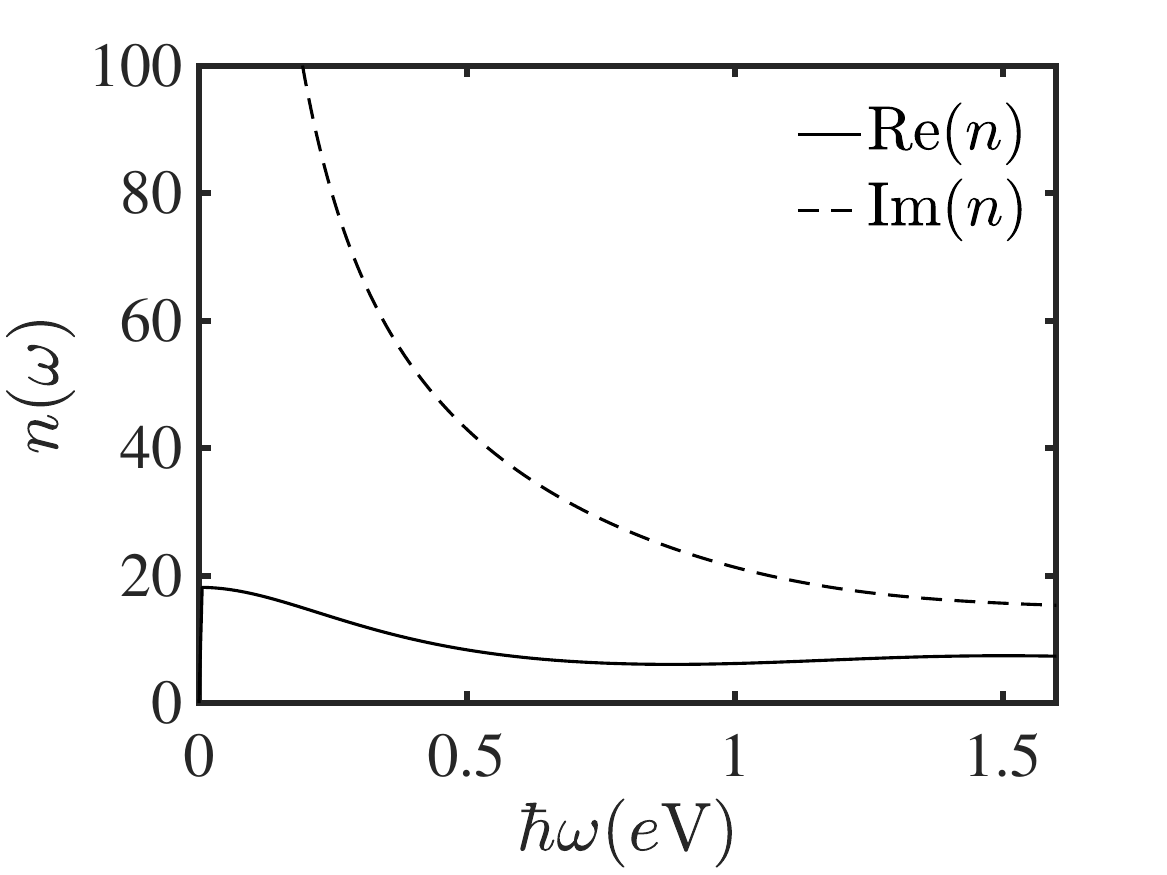}
            	\caption{Real (solid) and imaginary (dashed) parts of the frequency-dependent complex refractory index within the frequency region $\hbar\omega \in [0, 1.6]$ eV. }
            	\label{fig:refractory}
            \end{figure}

        	Following Ref.~\onlinecite{LZ2021} we take $d=\SI{9}{\angstrom}$, and following Ref.~\onlinecite{SZX2021} we use a Drude-Lorentz model to estimate the permeability tensor:
        	\begin{align}
        	\epsilon(\omega) = \epsilon_{\infty}-\sum_{k}\frac{\omega^2_{p,k}}{\omega^2+i\omega/\tau_k}+\sum_{j}\frac{\omega^2_{p,j}}{\omega^2_{0,j}-\omega^2-i\omega\gamma_j},
        	\end{align}
            where $\epsilon_{\infty}=10$. In the second term, $\omega_{p,k}$ is the plasma frequency, and $\tau_k$ is the lifetime of the quasiparticles of the $k$-th Drude oscillator. In the third term, $\omega_{0,j}$, $\gamma_j$, $\omega_{p,j}$ correspond to the resonance frequency, linewidth, and the plasma frequency of the $j$th Lorentz oscillator, respectively. All parameters are given in Table \ref{drude} and Table~\ref{lorentz}, which can be extracted from Ref.~\onlinecite{SZX2021}. With these parameters, we plot the frequency-dependent complex refractory index $n(\omega)$  as shown in Fig.~\ref{fig:refractory}. And then we obtain $\theta_{K}(\omega)$ using Eq.~\eqref{kerr_S}. We focus in particular on two frequencies $\hbar\omega=0.8~e\mathrm{V}$ and $1.55~e\mathrm{V}$ used in existing Kerr measurements. At $\hbar\omega=0.8~e\mathrm{V}$, $\epsilon=-692.84+333.14i$ and $n=6.16+27.03i$. At $\hbar\omega=1.55~e\mathrm{V}$, $\epsilon=-185.27+229.88i$ and $n=7.42+15.5i$.
            
        .


\begin{thebibliography}{99}
\bibitem{footnote} Here, we ignore the modulations to the real part of the electron hoppings on the bonds due to the CDW order, $\lambda^\prime$. Including such modulations will at best lead to $\mathcal{O}(\lambda^\prime/t)$ correction to the anomalous Hall conductivity.
\end{thebibliography}
\end{document}